          [2001/04/25 1.1 (PWD)]
\documentclass[a4paper,twocolumn]{esapub} 
\usepackage{times}
\usepackage{natbib}
\usepackage{epsfig}

\def \sig      {$\sigma$}
\def \gray     {$\gamma$-ray}

\title{
Multifrequency Observations of the Blazar 3C~279 in January 2006
}
\author[1]{W.~Collmar}
\author[2]{M.~B\"ottcher}
\author[3]{T.~Krichbaum}
\author[1]{E.~Bottacini}
\author[1]{V.~Burwitz}
\author[4]{A.~Cucchiara}
\author[4]{D.~Grupe}
\author[5]{M.~Gurwell}
\author[6]{P.~Kretschmar}
\author[7]{K.~Pottschmidt}
\author[8]{M.~Bremer}
\author[9]{S.~Leon}
\author[9]{H.~Ungerechts}
\author[10]{P.~Giommi}
\author[10]{M.~Capalbi}
\author[ ]{the WEBT collaboration}

\affil[1]{Max-Planck-Institut f\"ur extraterrestrische Physik, P.O. Box 1312, 85741 Garching, Germany}
\affil[2]{Department of Physics and Astronomy, Ohio University, Athens, OH 45701, USA }
\affil[3]{Max-Planck-Institut f\"ur Radioastronomie, Auf dem H\"ugel 69, 53121 Bonn, Germany}
\affil[4]{Pennsylvania State University, 525 Davey Lab, University Park, PA 16802, USA}
\affil[5]{Harvard-Smithsonian Center for Astrophysics, 60 Garden Street, Cambridge, MA 02138, USA}
\affil[6]{European Space Astronomy Centre (ESAC), European Space Agency, PO Box 50727, 28080, Madrid, Spain}
\affil[7]{CASS, Code 0424, University of California at San Diego, La Jolla, CA 92093, USA}
\affil[8]{IRAM, Avenida Divina Pastora 7, Local 20, E-18012 Granada, Spain}
\affil[9]{IRAM, 300 rue de la Piscine, Domaine Universitaire, 38406 Saint Martin d'Hères, France}
\affil[10]{ASI Science Data Center, ASDC c/o ESRIN, via G. Galilei, 00044 Frascati, Italy}

\begin{document}


\maketitle

\begin{abstract}
We report first results of a multifrequency campaign from radio to hard X-ray 
energies of the prominent \gray\ blazar 3C~279, which was organised around 
an INTEGRAL ToO observation in January 2006, and triggered on its 
optical state. The variable blazar was observed at an intermediate 
optical state, and a well-covered multifrequency spectrum from radio
to hard X-ray energies could be derived. The SED shows the typical
two-hump shape, the signature  of non-thermal synchrotron and 
inverse-Compton (IC) emission from a
relativistic jet. By the significant exposure times of INTEGRAL and Chandra, 
the IC spectrum (0.3 - 100 keV) was most accurately measured, showing 
-- for the first time -- a possible bending. 
A comparison of this 2006 SED to the one observed in 2003, 
also centered on an INTEGRAL observation, during an optical low-state,
reveals the surprising fact that -- despite a significant change at the
high-energy synchrotron emission (near-IR/optical/UV) -- the rest of the
SED remains unchanged. In particular, the low-energy IC emission 
(X- and hard X-ray energies) remains the same as in 2003, 
proving that the two emission components do not vary simultaneously,
and provides strong constraints on the modelling of the overall
emission of 3C~279.

\end{abstract}

\section{Introduction}

The discovery by the experiments aboard the Compton Gamma-Ray Observatory
(CGRO) that blazars can radiate a large -- sometimes even the major --
fraction of their luminosity at \gray\ energies marked a milestone
in our knowledge on blazars.
During the CGRO mission about 90 blazars were detected by the different 
CGRO experiments at \gray\ energies, the majority 
by the EGRET experiment at energies above $\sim$100~MeV
\citep{Hartman99}. 3C~279, an optically violently variable (OVV) quasar
located at a redshift of 0.538, is one of the most prominent
representatives of these sources.  
The source shows rapid variability in all
wavelength bands, polarized emission in radio and optical,
superluminal motion, and a compact radio core with a flat radio spectrum. 
These properties put the quasar 3C~279 into the blazar sub-class 
of AGN. According to the unified model of Active Galactic Nuclei (AGN),
blazars are sources which expel jets close to our line-of-sight.

3C~279 was already detected by INTEGRAL in June 2003 \citep{Collmar04}.
Those high-energy observations were supplemented 
in X-rays by a short (5 ksec) Chandra  pointing, and by ground based
monitoring from radio to optical bands. Since the blazar was found
in 2003 at the faintest optical brightness (optical R-band: $\sim$17~mag)
of the last 10 years, roughly 5 mag fainter than the maximum, and about 2.5 to 3 mag
fainter than average, a simultaneous spectral energy distribution (SED) of an 
exceptional optical low-state could be compiled \citep{Collmar04}.
In order to measure emission changes as function of optical brigthness,
we proposed for an INTEGRAL ToO observation during an optical high state
(criterion: optical R-band brigther than 14.5~mag). In January 2006, 
the trigger citerion was met, and the INTEGRAL observations together with 
supplementing multifrequency observations were carried out. This 
campaign resulted in a well covered SED from radio to hard X-ray energies.

In this paper we present first results of this
2006 multiwavelength campaign on 3C~279.
Because of the page restrictions, we concentrate on presenting
the main observational results, focussing on the observed SED 
and its comparison to the one of 2003. 
A more detailed presentation, including a discussion on the 
scientific implications of the new results, will be given 
in a later paper (Collmar et al., in prep.; including also all participating 
individual WEBT (Whole Earth Blazar Telescope) collaborators in the author list).
In addition, the results on variability analyses and time correlations 
of the different wavelength bands will be given elsewhere (B\"ottcher 
et al., in prep.).  

\begin{figure}[th]
\centering
\epsfig{figure=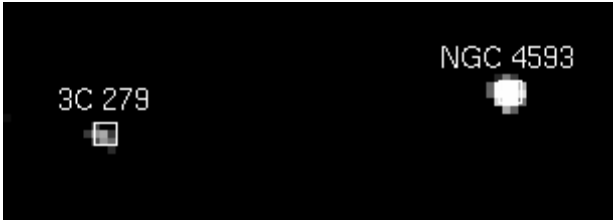,width=8.0cm,clip=}
\epsfig{figure=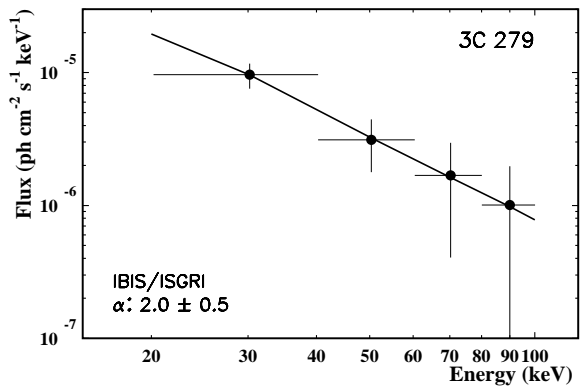,width=8.1cm,clip=}
\caption{
Top: The ISGRI image shows a 6.4\sig\ detection of 3C~279 in the  
20-60~keV band. In addition, the Seyfert galaxy 
NGC 4593 is even more clearly detected.
\newline Bottom: The ISGRI hard X-ray spectrum between 20 and 100 keV
is shown, together with the best-fit power-law shape.  
\label{fig:1}
}
\end{figure}
    
\section{Observations}

After the trigger criterion, 3C~279 brigther in the optical R-band 
than 14.5~mag, was met in early January 2006, we requested the granted ToO 
observations with INTEGRAL and Chandra. These INTEGRAL observations were carried out 
for a total of 511~ks between January 13 and 22, 2006 in several pointings, 
and the Chandra observations on January 17, 2006 for 30~ks.
Centered on these high-energy observations, we started a multifrequency campaign,
by initating simultaneous measurements in radio and mm-bands 
(Effelsberg and Mets\"ahovi radio telescopes, IRAM (Plateau de Bure and Pico Veleta), 
and the Smithsonian Submillimeter Array at Mauna Kea), in near-IR and optical bands 
by a WEBT campaign, and additional 
high-energy coverage (UV, X-rays) by Swift. 
This campaign resulted in a wealth of data, which -- although not all have been 
reduced and analysed yet -- yielded a well-covered SED of 3C~279 from radio to 
hard X-ray energies. In particular, by the participation of satellite experiments 
Swift, Chandra, and INTEGRAL, all with significant exposure times, the best-ever
spectral coverage in UV-, X-ray -, and hard X-ray energies was derived. 

The data analysis revealed, that 3C~279 was observed 
during an intermediate optical state.
With an R-band magnitude of about 15~mag, the source was $\sim$2~mag brighter
than during the low-state observations in the 2003 campaign.

\section{Results}

\subsection{High-Energy Observations}

The IBIS/ISGRI experiment aboard INTEGRAL detected the blazar at 
energies between 20 and 100 keV with a significance of $\sim$7.5$\sigma$
during the total 511~ks observation. The ISGRI image (Fig.~\ref{fig:1})
of the 20-60~keV band shows a 6.4$\sigma$ detection of 3C~279. 
Other Virgo region sources are also detected in these observations, 
like the prominent quasar 3C~273 and the Seyfert galaxy NGC~4593 for example. 
The blazar was measured by ISGRI at a surprisingly low flux level of 
$(2.53\pm0.52) \times 10^{-4}$ ph cm$^{-2}$ s$^{-1}$.  
Spectral analysis between 20 and 100 keV, assuming a power-law shape,
yields a weakly determined average spectral shape (Fig.~1) of photon index 
2.0$\pm$0.5 (1$\sigma$).   
Due to annealing, no INTEGRAL SPI data are available.
JEM-X did not detect 3C~279. Upper limits were derived for two
energy bands (5-10 and 10-20 keV) based on the mosaic images from
these observations.

\begin{figure}[th]
\centering
\epsfig{figure=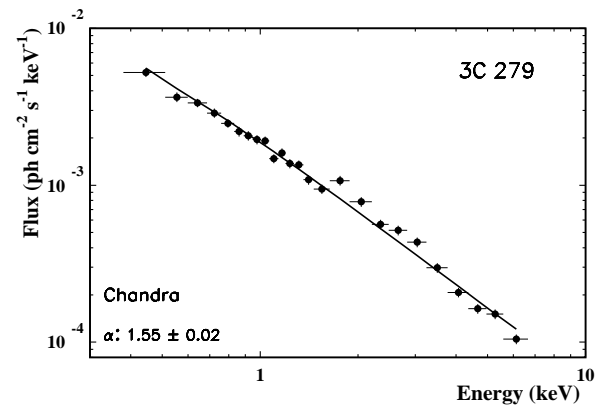,width=8.1cm,clip=}
\caption{
The Chandra 0.3 - 7 keV X-ray spectrum is shown,
 together with the best-fit power-law shape.
\label{fig:2}
}
\end{figure}

Chandra observed the blazar within the INTEGRAL 
observational period and simultaneously to INTEGRAL. To avoid pile-up 
in the Chandra detectors of the assumed strong X-ray source, 
the LETG-ACIS-S mode was used. 3C~279 is significantly detected by Chandra, 
and a well-determined X-ray spectrum between 0.3 and 7 keV could be
measured (Fig.~\ref{fig:2}). Assuming the canonical power-law shape at X-ray energies,
a spectral photon index of 1.55$\pm$0.02 was derived. 
The spectral analysis however, indicates  a trend for a
spectral bending from a harder to a softer spectrum towards higher 
energies. In particular, no soft excess was found, which would have 
indicated a contribution of the synchrotron component to the soft X-ray emission. 
Chandra did not observe any variability of 3C~279 during the uninterrupted 
source pointing of $\sim$8~hours.

\begin{figure}[bh]
\centering
\epsfig{figure=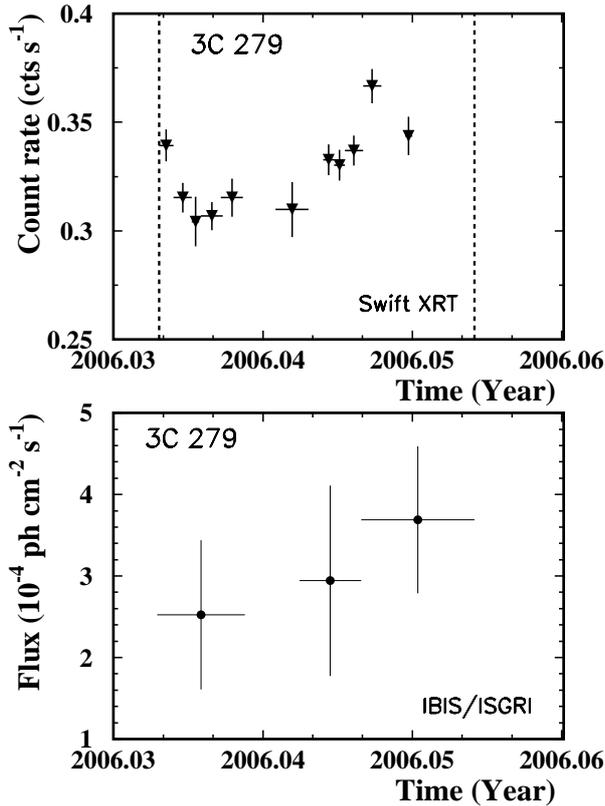,width=8.1cm,clip=}
\caption{
The variability at X-ray (top: Swift/XRT 0.2-10~keV) and hard X-ray 
(bottom: INTEGRAL/ISGRI: 20-60 keV) energies of 3C~279 during the campaign.
The dashed lines indicate the INTEGRAL observational period.
\label{fig:3}
}
\end{figure}

These high-energy INTEGRAL and Chandra observations were valuably 
supplemented by a series of Swift pointings towards 3C~279. 
Between January 13 and 20, 2006, Swift observed the blazar regularly, 
thereby providing important additional information on the source, 
like an X-ray light curve as measured by the XRT, and the UV fluxes
as measured by the UVOT. The Swift XRT (0.2-10 keV) X-ray light curve,
together with the IBIS/ISGRI (20-60 keV) hard X-ray light curve 
is shown in Fig.~\ref{fig:3}. While Swift observed significant
flux variability in X-rays, which can be utilized for correlations studies
to other bands, ISGRI -- measuring with less statistical significance --
can not detect significant variability, although there is a possible trend for a
 brigthening towards the end of its observational period. 
The Swift XRT spectra of individual pointings agree 
completely in spectral index with the one measured by Chandra. 
The Swift UV measurements provided additional coverage of the SED (Fig.~6).

\begin{figure}[tb]
\centering
\epsfig{figure=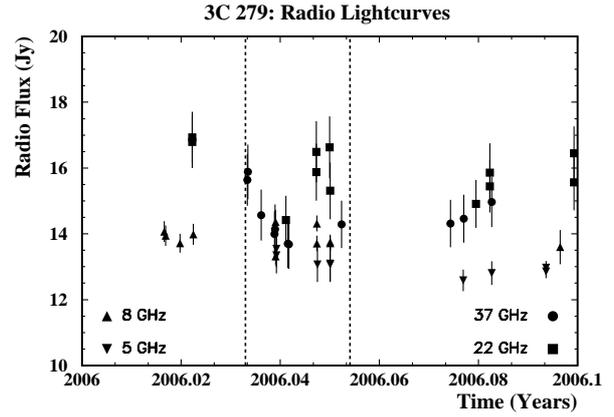,width=8.1cm,clip=}
\caption{
Radio light curves from monitoring observations 
at or around the INTEGRAL observational period, which is
indicated by the vertical dashed lines.
\label{fig:4}
}
\end{figure}

\begin{figure}[h]
\centering
\epsfig{figure=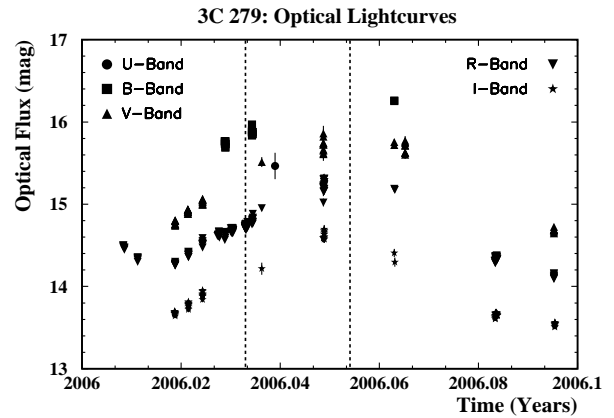,width=8.1cm,clip=}
\caption{Optical monitoring by the WEBT-campaign at or around the 
high-energy INTEGRAL observations, indicated by the vertical dashed lines.   
\label{fig:5}
}
\end{figure}

\subsection{Low-Energy Observations}

Since the start of INTEGRAL AO-3, 3C~279 
was regularly observed in the optical
-- in particular during visibility periods of INTEGRAL --
in order to keep track of its brigthness state.
Early January 2006, the blazar exceeded the trigger level for the satellite
observations, which then were initiated.
In order to have good ground-based coverage, a WEBT 
campaign was requested simultaneous to the high-energy observations. 
This campaign resulted in broadband (radio to optical)
ground-based monitoring of the blazar. Subsequently light curves
(for correlation studies) and average flux measurements (SED) in different 
wavelength bands could be compiled. The optical monitoring revealed, that in 
the R-band the blazar brightened early January up to $\sim$14.2~mag on January 5, 
and then continously faded during the next 3 weeks, being at
$\sim$14.7~mag at the beginning and $\sim$15.2~mag at the end of the 
INTEGRAL observations. This may indicate a continous cooling following an
optical flare. The radio and optical light curves of the camapign are given 
in Figs.~\ref{fig:4} and \ref{fig:5}. 
In this page-limited proceedings paper, we shall concentrate on the measured 
SED. A more detailed analysis of the broadband variability, including inter-band 
cross correlations and their physical interpretation will be presented
in a forthcoming paper.

\begin{figure}[h]
\centering
\vspace*{-0.5cm}
\epsfig{figure=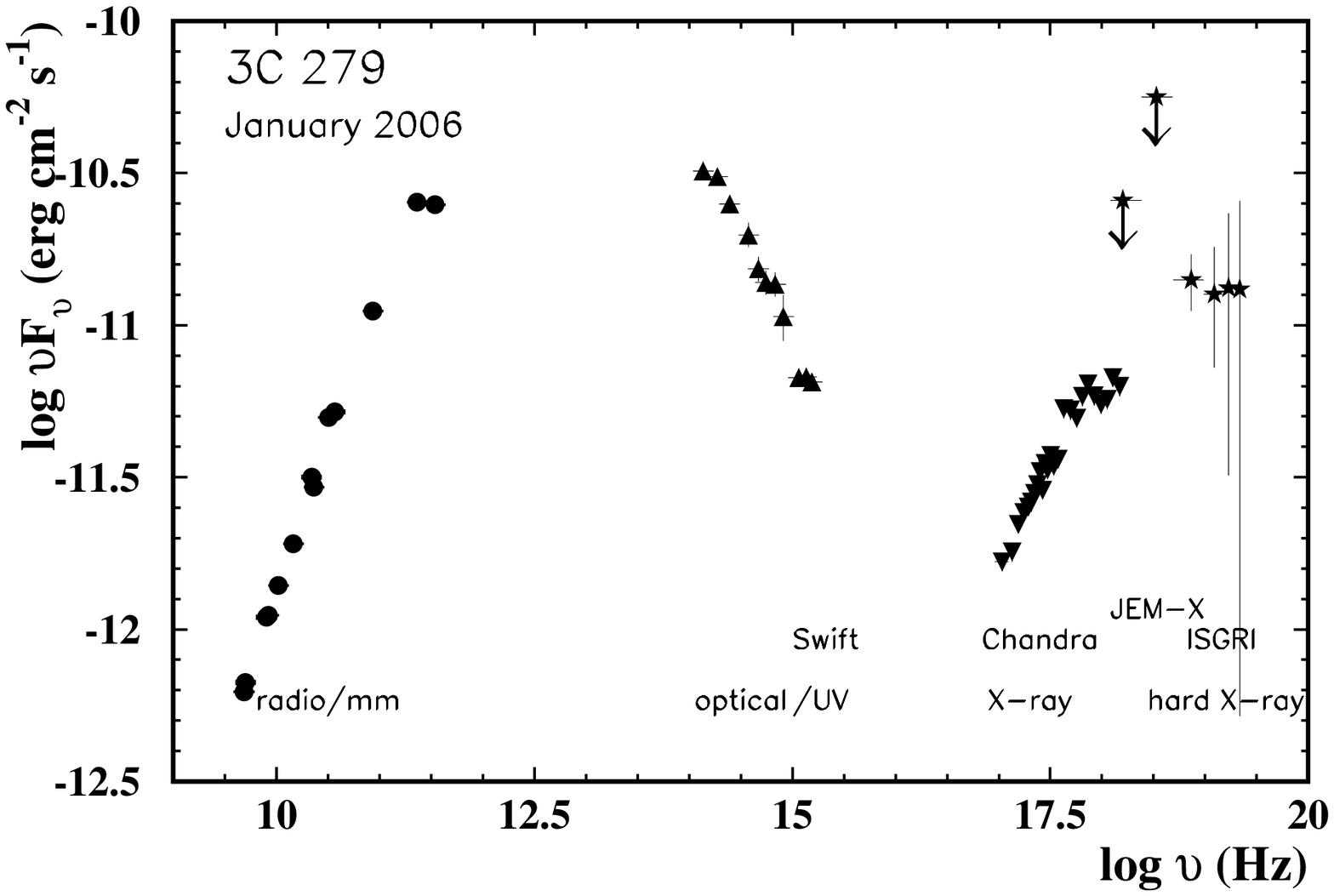,width=8.4cm,clip=}
\epsfig{figure=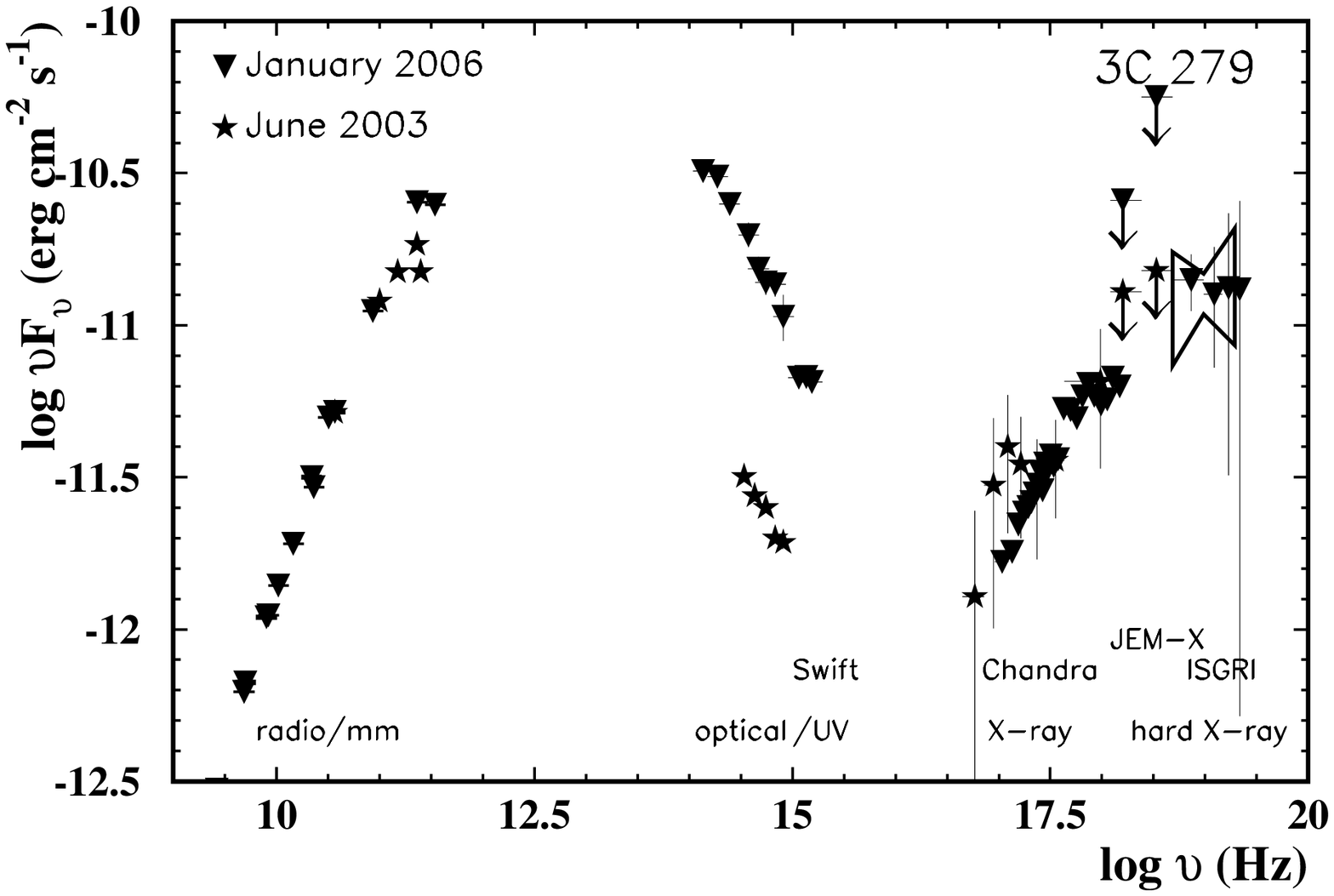,width=8.4cm,clip=}
\caption{
Top: The 3C~279 SED as measured during this campaign in 
January 2006. The spectrum, well covered from radio to hard X-ray energies, 
shows the typical two-hump shape, with an unobserved crossover
between UV (measured by the Swift UVOT) and soft X-rays (0.3~keV).
Flux points are shown with 1\sig\ error bars. 
Upper limits are 2\sig. At low energies the error bars are smaller
than the symbols. 
\newline Bottom: Comparison of the January 2006 (intermediate optical state)
multiwavelength spectrum to the low-state multifrequency spectrum of 
June 2003 [1].
While the high-energy part of the synchrotron peak (near-IR, optical, UV)
is significantly different, the observed low-energy (0.3 - 100 keV) 
part of the IC peak is unchanged.
\label{fig:6}
}
\end{figure}

\subsection{Multiwavelength Results}

Fig.~\ref{fig:6} (top) shows the contemporaneous SED of 3C~279 during January 2006.
The SED, although not strictly simultaneous, is compiled only from data, 
measured within the INTEGRAL observational period. For example, the IBIS/ISGRI 
spectrum is averaged over the complete INTEGRAL period, while the Chandra spectrum 
is -- of course -- from the 8-hour observation on January 17. 
The SED is well covered from radio to hard X-ray energies and
shows the typical two-hump shape, which is belived to be synchrotron 
at the lower- and IC emission at the higher energies.
The synchrotron peak, located probably somewhere at IR energies, is not observed. 
The crossover point of the synchrotron and IC emissions is also not observed, 
because Chandra did not find an excess at soft X-rays, which are already
located at the rising IC branch. 
The SED indicates a deep minimum between the two branches.
The significant exposures of Chandra and ISGRI yielded the yet most accurate 
spectral shape measurement of the rising IC emission of 3C~279. A spectral bending 
from a harder towards a softer spectrum is indicated. 
A modelling of this spectrum by assuming a leptonic emission model 
is in progress (Collmar et al., in prep.).

In Fig.~\ref{fig:6} (bottom), this January 2006 SED is compared to 
the optical low-state one as measured in June 2003 \citep{Collmar04}.
The observations in 2006 were selected such (ToO) that the blazar was 
significantly brigther in the optical than  in 2003.
The goal was to observe how the SED changes with a changing optical 
flux. The surprising result is, that despite a significant
change in near-IR, optical, and UV (the high-energy part
of the synchrotron emission), the rest of the SED is unchanged.
In particular, the low-energy part (0.3 - $\sim$100~keV) of the 
IC emission remains completely unchanged. This result shows that there
is no -- at least not simultaneous -- correlation between these two bands. 
Whether the excess synchrotron photons have a counterpart 
at higher IC energies (MeV, GeV energies)
and/or whether energy-dependent time delays occur during the cooling off of 
a flare, e.g. higher-energy photons are emitted earlier than lower-energy ones, 
cannot be resolved by this SED comparison. Detailed inter-band variability studies
(possible by the wealth of data), 
and annother such measurement during the upcoming 
GLAST era, when also the MeV/GeV part of the IC emission will be observed, 
will shed light on these possibilities.

\end{document}